\newcommand{\beq}{\begin{equation}}
\newcommand{\eeq}{\end{equation}}

\newcommand{\bear}{\begin{eqnarray}}
\newcommand{\eear}{\end{eqnarray}}

\newcommand{\tn}{\textnormal}

\documentclass[a4paper]{jpconf}
\usepackage{graphicx}
\usepackage{enumerate}
\usepackage{pifont}
\usepackage{amssymb}
\usepackage{bbding}
\usepackage{lineno}
\begin{document}
\title{Lessons from the operation of the `Penning-Fluorescent' TPC and prospects}

\author{Diego Gonzalez-Diaz, F. Aznar, J. Castel, S. Cebri\'an, T. Dafni, J. A.~Garc\'ia, J. G. Garza, H. G\'omez, D. C. Herrera, F. J. Iguaz, I. G. Irastorza, A. Lagraba, G. Luz\'on, A. Rodr\'\i guez, E. Ruiz-Choliz, L. Segui, A. Tom\'as}
\address{Laboratorio de F\'isica Nuclear y Astropart\'iculas, Universidad de Zaragoza, Zaragoza, Spain}
\ead{diegogon@cern.ch}

\author{E. Ferrer-Ribas, I. Giomataris}
\address{IRFU, Centre d'\'Etudes Nucl\'eaires (CEA), Saclay, France}

\begin{abstract}
We have recently reported the development of a new type of high-pressure Xenon time projection chamber operated with an ultra-low diffusion mixture and that simultaneously displays Penning effect and fluorescence in the near-visible region (300 nm). The concept, dubbed `Penning-Fluorescent' TPC, allows the simultaneous reconstruction of primary charge and scintillation with high topological and calorimetric fidelity.
\end{abstract}

\section{Introduction and main results in charge mode}
\begin{figure}[ht!!!]
 \centering
 \includegraphics*[width=7.8cm]{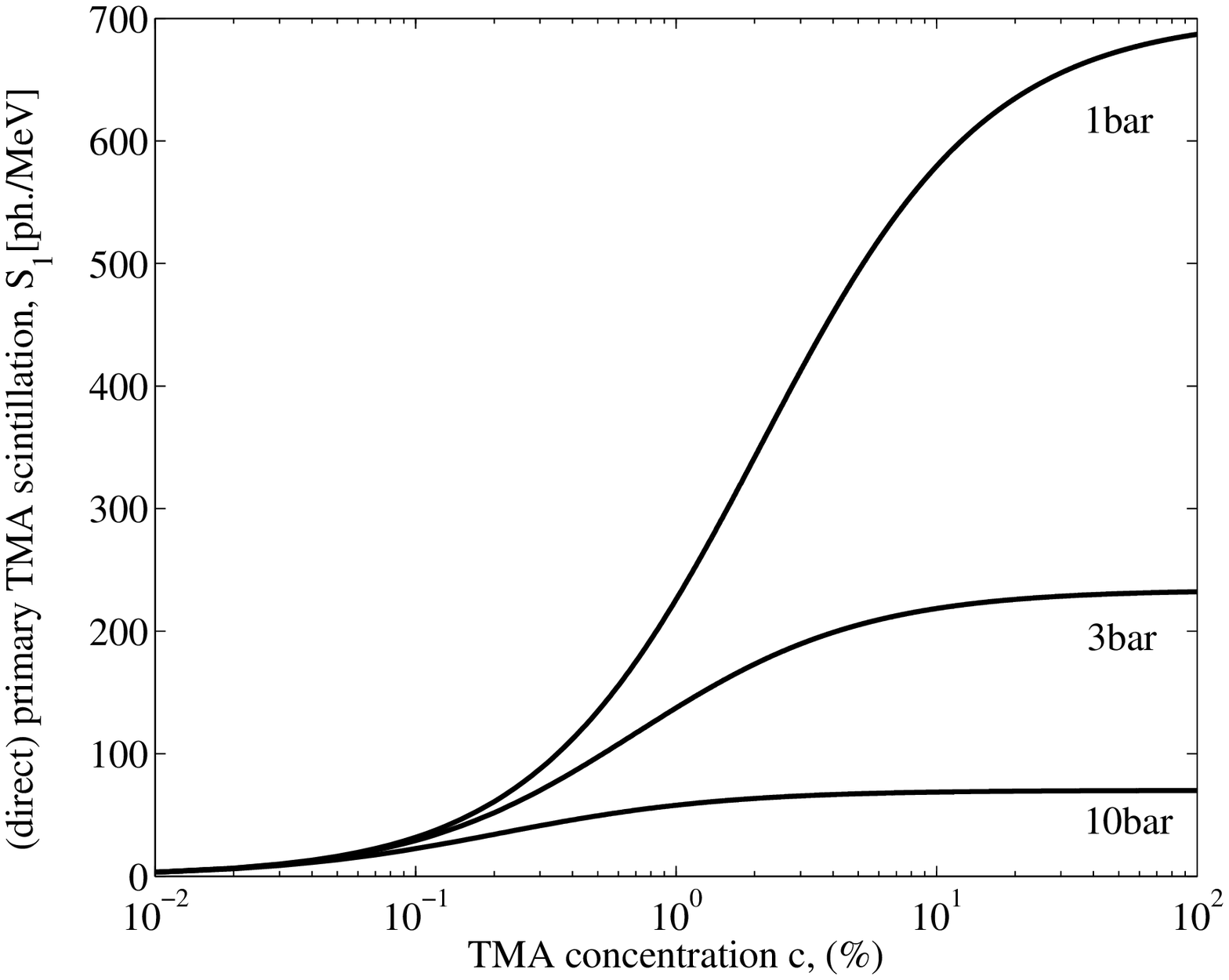}
 \includegraphics*[width=7.8cm]{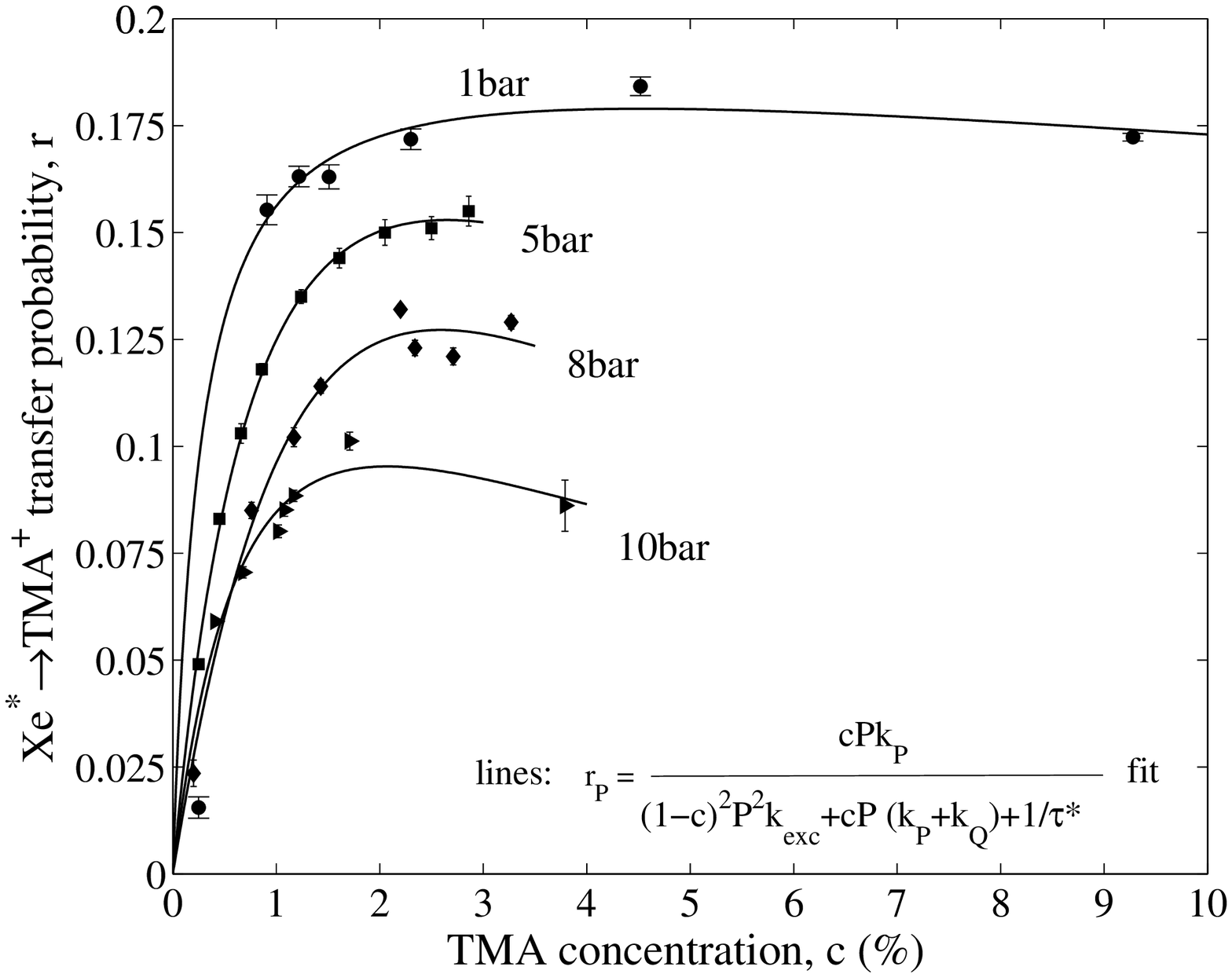}
 \caption{Left: estimated primary scintillation yields (S$_1$) in the TMA fluorescence band (300 nm) for Xe-TMA mixtures under the conservative assumption that only direct TMA excitation produces observable photons. Right: Penning transfer rates obtained from a first-order modelling of the gain observed in a Micromegas charge amplification device, assimilated to a parallel plate chamber. A detailed 3D field calculation yields maximum transfer rates that are higher by up to 50\%, and will be published elsewhere \cite{Eli}.}
 \label{S1yields}
 \end{figure}

\begin{figure}[ht!!!]
\centering
\includegraphics*[width=4.3cm]{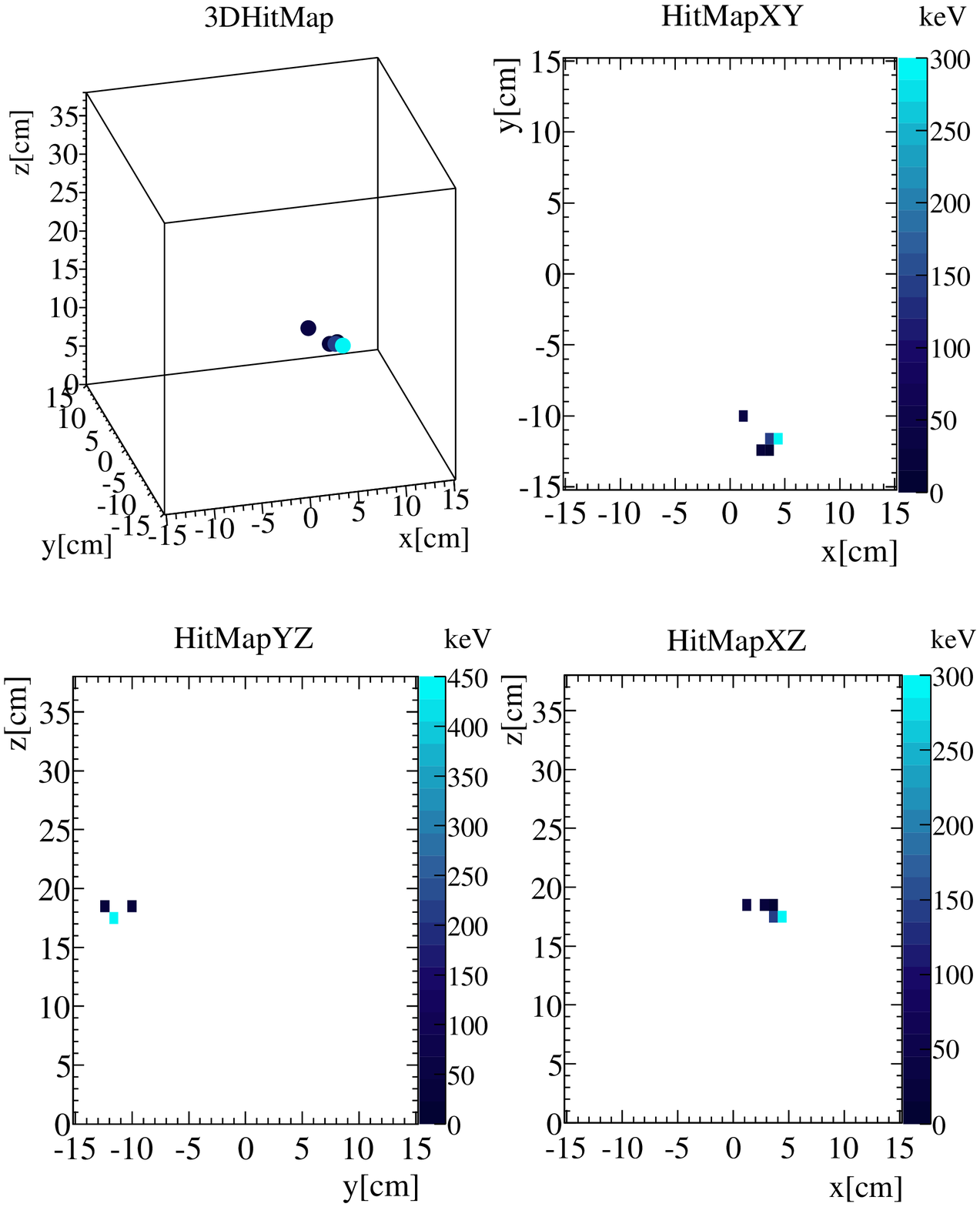}
\includegraphics*[width=4.45cm]{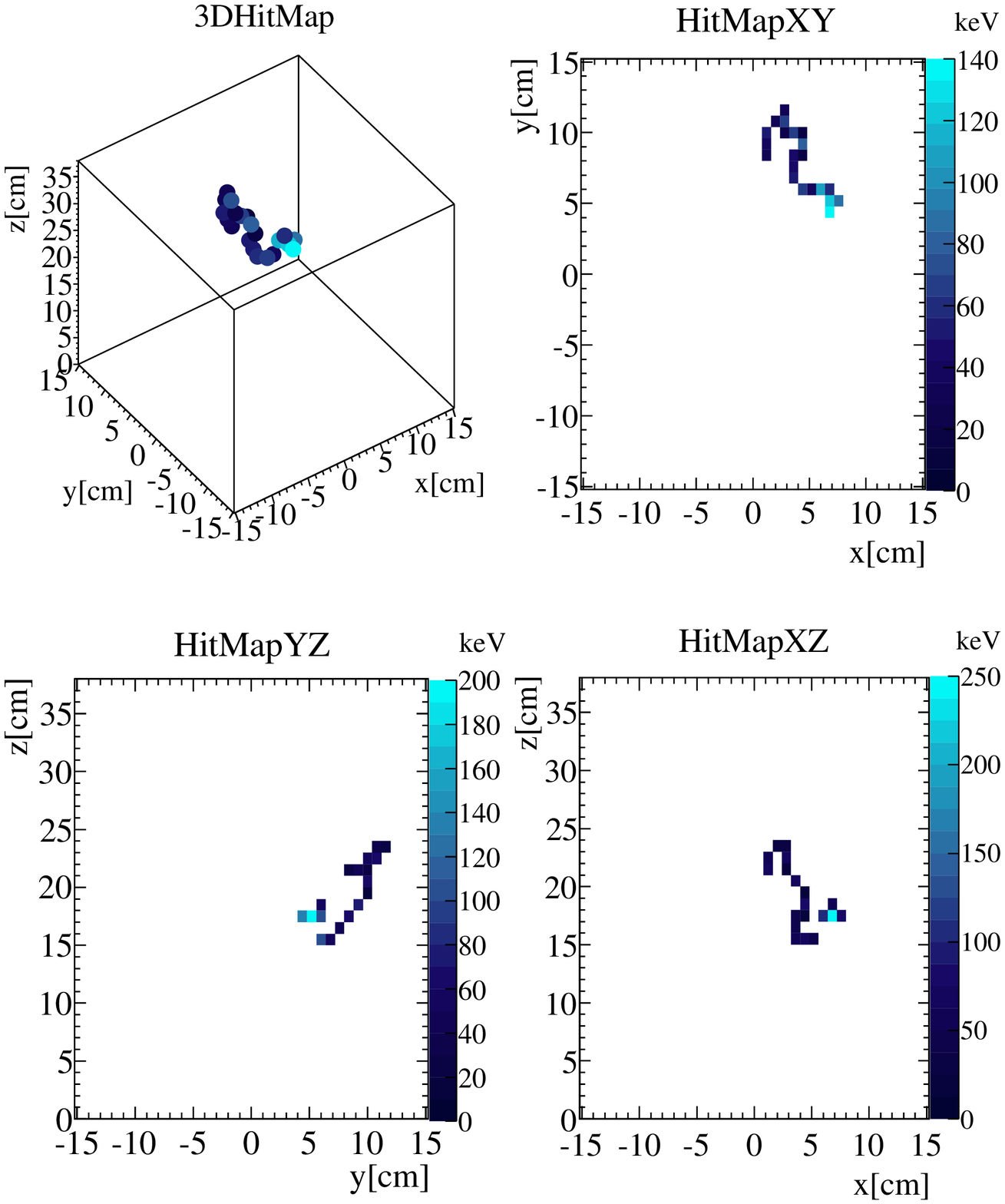}

\includegraphics*[width=4.4cm]{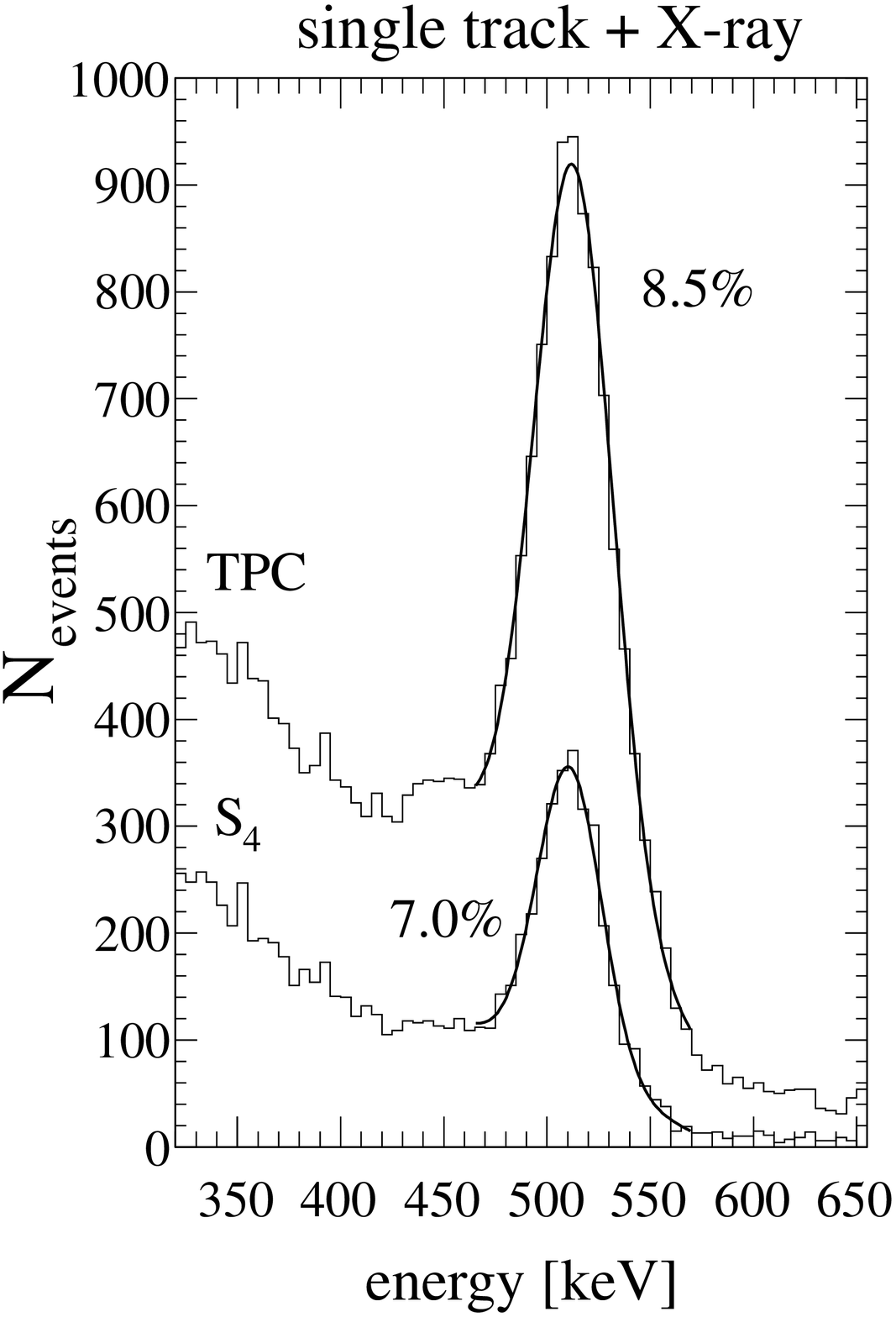}
\includegraphics*[width=4.4cm]{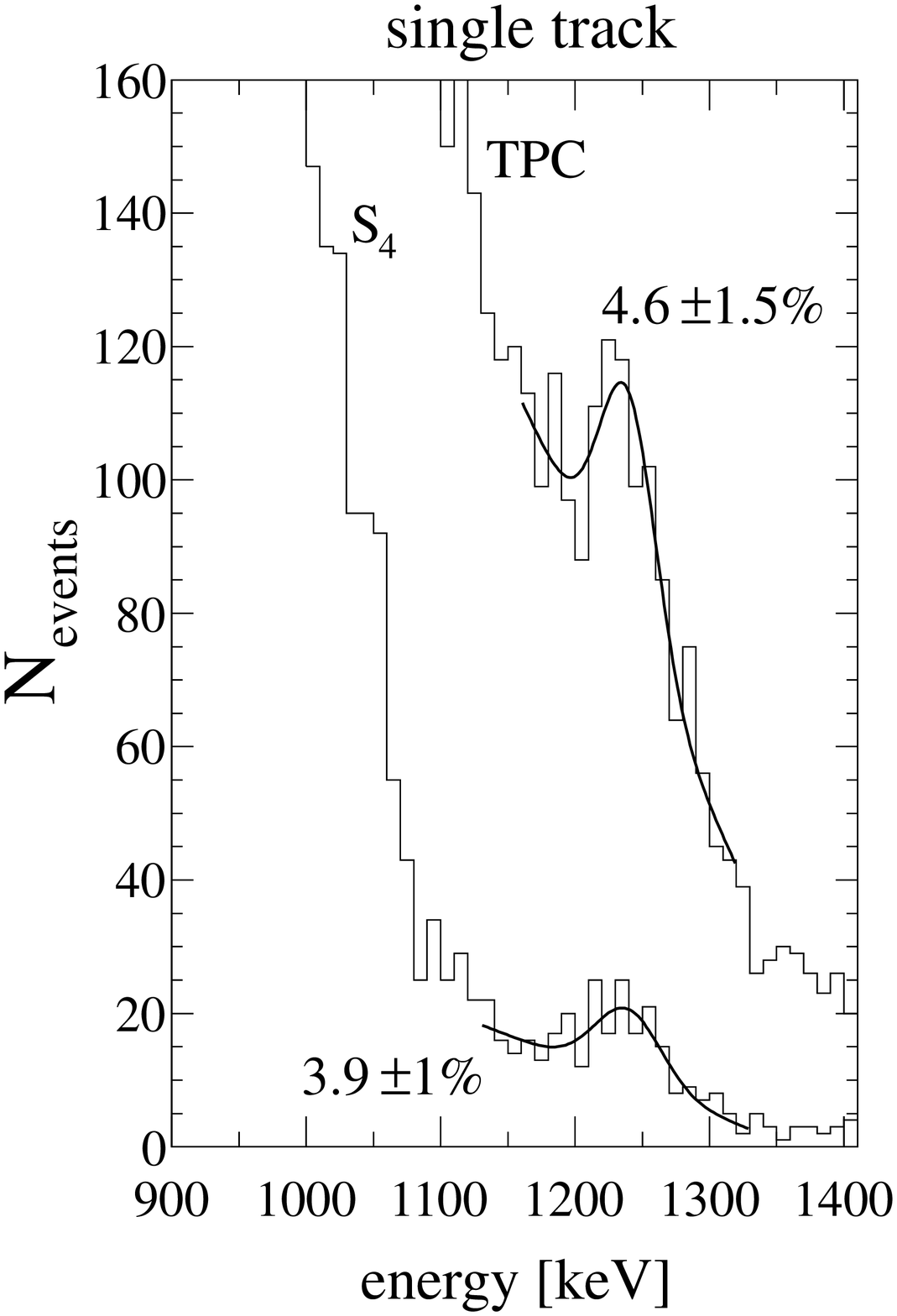}
\caption{Up: 3D reconstruction and 2D projections of typical 0.511 MeV (left) and 1.275 MeV $\simeq Q_{\beta\beta}/2$ (right) tracks, obtained in the fiducial volume (1.1kg) of the large technological demonstrator NEXT-MM, shown with a voxelization of 8 mm $\times$ 8 mm $\times$ 10 mm ($x$, $y$, $z$). Down: calorimetric response for each case, indicating FWHM values. The energy resolution for 30 keV X-rays is 14.6\%(FWHM).}
\label{511}
\end{figure}

The possibility of using `Penning-Fluorescent' gas mixtures was addressed as early as at least 1981, with the seminal work of A. Policarpo on light multiplication \cite{Poli}, and coetaneous to the invention of the TPC by  D. Nygren and J. Marx \cite{TPC}. The possibility of the confluence of the two techniques was brought forward recently, by Nygren himself \cite{DavePF}, in the context of novel gas mixtures for $\beta\beta0\nu$-searches, for which the TPC technology must be desirably pushed beyond the current technological limits both in the calorimetric and topological front. We have extensively studied the possibility of using Xenon/trimethylamine Penning-Fluorescent mixtures in a series of papers \cite{XeTMA, Diana2, RecoProc, HectorTMA, DiegoTMA} culminating in \cite{DiegoTMALast}, and the main lessons there extracted are elaborated further in this proceeding. A summary of the impact of the TMA addition on the primary charge and scintillation of gaseous Xenon is summarized in Fig.\ref{S1yields}, while the topological and calorimetric properties obtained inside a 1.1 kg fiducial mass (as recorded with microbulk Micromegas) are given in Fig.\ref{511}. An extensive description of the chamber performance for the reconstruction of $\gamma$-tracks in charge mode can be found in \cite{DiegoTMALast}, and for the light properties the reader is referred to the work of Y. Nakajima also presented at this conference \cite{Yasu}. A microscopic modelling of the Micromegas response is in preparation \cite{Eli}.

Importantly, it must be noted at the outset that the term `Penning-Fluorescent' refers here to wavelength-shifted fluorescence, i.e., to scintillation with a characteristic spectrum centered around the visible range (hence, noble gas VUV-fluorescence must be necessarily quenched by the mixture too). In this shifted band, the parasitic photo-effect at metallic surfaces as well as photon transmission become extremely benign, facilitating both the detection of the primary scintillation with conventional PMTs and charge detection with micro-pattern gaseous detectors.

\subsection{Fluorescence}
The TMA fluorescence spectrum is centered at around 300 nm and ends at about the work function of copper, while TMA itself is strongly opaque to Xenon light. A lower bound to the primary scintillation expected from a Xenon-TMA mixture can be derived from studies at low pressure, and clearly requires confirmation at the pressures of interest for $\beta\beta0\nu$: omitting conservatively the presence of Xe$^*$, Xe$_2^*\rightarrow$TMA$^*$(3s) transfers and hence assuming only the processes of direct TMA excitation/scintillation and self-quenching \cite{Cureton} and neglecting thus the quenching of TMA excited states by Xenon species (following \cite{ObiXe}) such a bound results in:
\beq
S_1(TMA) = \frac{1}{W_{sc}} \frac{c}{\tau_{3s}(P \cdot c \cdot k_{SQ} + 1/\tau_{3s})}
\eeq
as a function of the concentration ($c$) and pressure ($P$).
Predicted $S_1$ yields from direct TMA excitation are shown for illustrative purposes in Fig. \ref{S1yields}, for 1, 3 and 10 bar. It has been assumed for TMA a typical value $W_{sc}=30\pm{10}$ eV ($\simeq 5.5 \times E_{3s}$), the measured lifetime of the 3s state ($\tau_{3s}=44$ ns) and the TMA self-quenching rate ($k_{SQ}= 1.08$ ns$^{-1}$ bar$^{-1}$). Additional scintillation hangs on the ability of TMA to quickly dissipate, collisionally, a fairly large amount of internal excess energy ($\sim 2$ eV) following the Xe$^*$, Xe$_2^*\rightarrow$TMA$^*$ transfers; relaxation to the 3s state may in this way happen while avoiding the dissociative pathways, \cite{ObiXe}. S$_1$ yields as those shown in Fig. \ref{S1yields}-left are not hopeless for $\beta\beta0\nu$-experiments or $\gamma$-detection in general but certainly require of dedicated detection schemes. A comprehensive experimental survey of the primary and secondary scintillation processes is currently under preparation and preliminary results can be found in \cite{Yasu}. Although that work points in fact to the smallness of the fluorescent transfer rate in Xe-TMA mixtures, the observed scintillation in the TMA band is compatible with the lower bounds given in Fig. \ref{S1yields}.

\subsection{Penning transfer rate}
To the extent that energy transfers occur with the Xe$^*$ states, Penning reactions are energetically viable:
\beq
\tn{Xe}^* + \tn{TMA} \rightarrow \tn{Xe} + \tn{TMA}^+ + \tn{e}^- \label{PenningEq}
\eeq
The presence of reactions of the type (\ref{PenningEq}) in Xe-TMA mixtures has been indirectly inferred from the strong gain increase at constant field observed in charge amplification structures (\cite{Ramsey}, \cite{XeTMA}), with an optimum admixture of about 1\% TMA for the case of Micromegas. In order to quantify the effect, we present here a simplified analysis performed by assimilating the Micromegas to a parallel plate configuration and using the microscopic code Magboltz for modelling the charge amplification process \cite{Magboltz}. An analogous procedure has been used in an earlier work for describing Argon Penning-mixtures in proportional counters \cite{Ozkan}.

Under the common simplifying assumption that the Penning transfer probability (hereafter $r_P$) is largely field-independent, a universal value can be extracted for it, as shown in Fig.\ref{S1yields}-right. A phenomenological description of the trends can be achieved in this case by considering excimer formation ($k_{exc}$), Penning ($k_{P}$) and quenching ($k_{Q}$) rates, and an effective life-time of the state from which Penning takes place ($\tau^*$), resulting in the relation:
\beq
r_P = \frac{c P k_P}{(1-c)^2P^2k_{exc} + c P (k_{P} + k_{Q}) + 1/\tau^*(P)}
\eeq
The observed reduction of transfer probability with pressure would seemingly indicate that Penning-transfer from the excimer states is disfavoured and so excimer formation effectively reduces the chances of the transfer to take place.

A more realistic analysis done by including a 3D field calculation will be presented elsewhere \cite{Eli}, although neither it substantially alters the trends nor the conclusions. The maximum transfer rate observed in such an analysis exceeds by 30-50\% (relative) the values given here, since the parallel plate approximation overestimates the amplification field.

\section{Discussion}

\subsection{$T_o$-less Micromegas-only TPC}

\begin{figure}[ht!!!]
\centering
\includegraphics*[width=11cm]{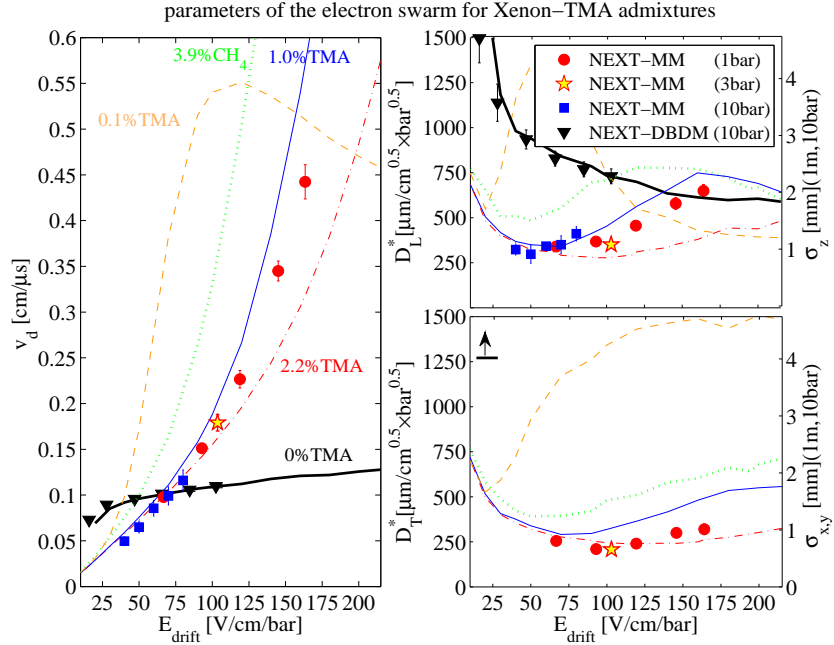}
\caption{Compilation of the main properties of high pressure Xenon TPCs for 1(1.0) 3(2.7) and 10(10.1) bar and different mixtures, together with simulation (Magboltz 10.0.1). Left: drift velocities. Right: longitudinal and transverse diffusion coefficients. Simulation results for Xe-CH$_4$ (Gotthard) and Xe-TMA in a 0.1\% admixture are also shown for illustration.}
\label{Swarm}
\end{figure}

The energy resolution obtained for 0.511 MeV tracks with Micromegas is at least $\times 2$ better than the one obtained earlier with wire chambers in \cite{Gotthard}, with Xe-TMA reducing electron diffusion in factors of $\simeq 2$ in all 3 space dimensions relative to, e.g., Xe-CH$_4$ (Fig. \ref{Swarm}) and operation taking place at twice the pressure, thus increasing event containment.
Since as of today microbulk Micromegas show good prospects for high radiopurity levels, it may seem that re-sizing the present concept `as is' could already improve on earlier limits obtained for $\beta\beta0\nu$-decay in high pressure Xenon in \cite{Gotthard}.
In this scenario, the optimum pressure is undoubtedly a crucial variable to be considered: the maximum gain reported is dramatically dependent on it, e.g. gains of 2000 were shown for small wafers at 5 bar in \cite{XeTMA} as compared to 400 at 10 bar, with 200 finally achieved in the system. Any future development would certainly have to address the interplay between signal to noise ratio (S/N) and operating pressure. These type of TPC configurations would constitute excellent X-ray or $\gamma$-ray polarimeters, provided the $T_o$ information is not essential \cite{HARPO}.\footnote{The possibility of obtaining the $T_o$ information from the physics of the electron diffusion (pulse width-$z$ correlation) should not be discarded and has been shown earlier at 1bar for localized X-ray irradiation \cite{DiegoTMA}. To the extent that enough S/N and sampling frequency are provided, and the information from displaced X-rays (when present) and/or the combined information from the whole track can be used, this method will find application. Since the low diffusion presumably required for the topological identification and suppression of $\gamma$ backgrounds is contrary to the high diffusion required for a achieving a strong width-$z$ correlation, the necessity of a dedicated optimization of the sampling frequency, signal shaping and buffer size is likely to ensue upon close examination of this possibility.}

\subsection{Micromegas TPC with $T_o$}
\label{TPCwithTo}
In $\beta\beta0\nu$ experiments the determination of the initial scintillation provides the time of the interaction ($T_o$) allowing event fiducialization and background suppression, as well as enhancing topological reconstruction due to the dependence of diffusion with $z$. Barium-tagging schemes usually take its knowledge for granted \cite{JJBaTa}, \cite{EXObata}, \cite{DaveBaTa}.

The level of direct TMA scintillation at 10 bar amounts to a modest yield of $S_1(\sim\!\!300\tn{nm}) = 58$ ph./MeV, increasing in approximate linear fashion with decreasing pressure.
As said, although additional energy transfer from excited Xenon species to the TMA fluorescent 3s-state may happen to some extent, it certainly does not represent a strong effect\cite{Yasu}.
Taking such a conservative scintillating figure and assuming a 25\% quantum efficiency, nearly a 0.3 solid angle coverage would be needed in order to collect 10 photons for
an event at the $Q_{\beta\beta}$ energy, that is certainly unrealistic to be done by sheer increase of the number of PMTs.
One may recycle for this purpose a recent idea brought forward in \cite{DaveReco}, benefiting from the fact that the TMA emission band matches the peak of the excitation spectrum for common commercial wavelength-shifting (wls) plastics.
Although challenging at present, this possibility would allow a drastic reduction in the number
of PMTs with additional benefits such as reduced cost and simplicity.

On the other hand such a TPC, with the interior of the field cage and the cathode back covered with wls-plastics would generically constitute an excellent X- and $\gamma$- ray detector with good prospects both as Compton camera and polarimeter. With optimized thickness, the surrounding plastics would simultaneously allow vetoing ionizing particles.

\subsection{Micromegas+EL TPC with $T_o$}

Despite the large improvement as compared to earlier approaches based on wires, the present
work shows that calorimetry with charge readouts still is, as of today, more prone to instrumental limitations than an electroluminescent (EL) approach, particularly at high pressure. This represents an important practical consideration for high precision discovery-type experiments. On the other hand, strategies concerning noise and cross-talk mitigation have been discussed in \cite{DiegoTMALast}, and a factor $\times 2$ improvement down to 1.5\% energy resolution (FWHM) at the energy scale of the $^{136}$Xe $Q_{\beta\beta}$ seems realistic without altering the Micromegas readout characteristics. The already modest rate of pixel damage (1\%/year) can be brought further down to the 1\%-level during realistic experiment live-times through the increase of the pixel granularity and the introduction of tighter quality control (QC) procedures.

Importantly, as compared to light-based readouts, charge amplification and readout with micro-pattern gaseous detectors can more easily approach a `true-3D' reconstruction (i.e., with the voxelization stemming from pixelization and signal sampling being at the scale at which the relevant track structures emerge: $\sim 1$mm$\times 1$mm$\times 1$mm at 10 bar \cite{DiegoTMALast}). So, based on the above paragraph it seems worth considering the possibility of a separate topological/calorimetric function, where both tasks can be optimized independently to a large degree. Such a speculative idea
could be implemented through a Micromegas readout placed immediately behind the EL region with its cathode serving
as the anode of the electroluminescence gap. Calorimetry can be performed with a PMT plane. This brings a number of considerations:

\begin{enumerate}
\item The large feedback and low gains reported in earlier measurements for Micromegas operated in pure Xenon \cite{PureXe} suggest that the use of a VUV-quencher is at least convenient (if not mandatory). At the m-scale Xe/TMA dumps virtually all VUV light, and yet it shows fluorescence in the TMA scintillating band, although special detection schemes are needed (subsection \ref{TPCwithTo}).
\item Quantitatively, it is unclear at the moment what the global effect of the addition of TMA in the calorimetric response is. In some admixtures it can conceivably reduce the electroluminescence threshold and Fano factor in virtue of its scintillating and Penning properties, however it may also increase the intrinsic fluctuations during the production of secondary light, specially if Penning from Xe$^*$ states would be unavoidable during the excitation of the TMA 3s (scintillating) states, hence yielding the characteristic non-linear light-multiplication trends reported in \cite{Yasu}. As a desirable advantage, the mixture will be transparent to the TMA scintillation at the scales of interest, and the nearly $4\pi$ coverage enforced by the necessary $S_1$ sensitivity will guarantee that only the intrinsic light fluctuations will be present, without additional fluctuations stemming from finite photon statistics.
\item Due to the relatively high EL-fields, electron transmission from the EL region to the Micromegas amplification holes will approach the optical transparency. This is as low as 12.5\% in the present case, although it easily doubles in a triangular pattern \cite{XeTMA}. Since, in this approach, Micromegas do not perform the calorimetric function this is not very critical, however the value should be increased to approach at least 50\% levels. Standard Micromegas manufactured in the bulk, provide easily higher levels of optical transparency.
\item Photo-cathode efficiencies of bare metals in the Xenon band are of the order of 10$^{-4}$ \cite{YannisPhoto}. Further to that, extraction efficiencies are extremely poor in Xenon ($\lesssim 10\%$), \cite{CollEffCoimbra}. For a typical optical gain of 200 ph./e/cm/bar this will reduce the contribution from parasitic photo-electrons to 1\%, and possibly much less since TMA fluorescence sits below the work function of copper and Xenon light will be strongly suppressed. So, stability of Micromegas on such a `photon-shower' seems to be reasonably possible.
\item At the Micromegas, light production during the multiplication process will add fluctuations to the reconstructed light. This needs to be assessed experimentally for Xe/TMA admixtures and cannot be discarded out of hand. There is however an important solid angle reduction in the light emerging from microbulk structures, that implies for pure Xenon a modest 20\% contribution to electroluminescence light, as reported in \cite{PureXe}. A compromise between high optical transmission and low scintillation from the Micromegas structure will be needed.
\item The Micromegas plane will be exposed to sparks from the EL-region, that will likely result in permanent damage, and potentially imply a catastrophic event. It requires special attention and its stability should be demonstrated.
\item From the above points it is clear that the incorporation of auxiliary micro-patterned structures for extraction from the EL-region and for pre-amplification (e.g., GEMs or grids) can easily alleviate the most stringent requirements, implying at the same time little impact on the tracking capabilities at the mm-scale.

\end{enumerate}

\section{Conclusions}
After a series of works performed over the last 3 years, TMA becomes established as an adequate additive for the construction of large Time Projection Chambers, with a number of remarkable properties that make it unique. It is a good Xenon VUV-quencher, yet demonstrating sufficient primary scintillation for MeV-scale track reconstruction and thus simultaneously allowing stable operation of charge amplification micro-pattern devices, by virtue of its wavelength-shifted emission at about 300 nm. It shows Penning effect, therefore enhancing the ionization response, reducing the field needed for gas amplification and potentially improving the intrinsic energy resolution both for primary and secondary ionization, if all spurious instrumental limitations are conveniently mitigated.
TMA can be counted amongst the strongest coolants employed in TPCs, comparable to CO$_2$, CF$_4$ or DME, and indeed the TPC here introduced shows the smallest electron diffusion of any TPC that is known to authors, at the scale of 1 mm per 1 m drift. TMA has a convenient vapor pressure, exceeding 1bar at ambient temperature, and it can be incorporated to re-circulating systems even in the presence of purifiers. Toxicity and flammability hazards benefit from its strong odour (easily detectable in ppb concentrations), hence posing yet stronger practical limits to the overall gas tightness, limits that were nevertheless comfortably achieved during operation of the TPC whose performance has been summarized here.

Although presently limited in the scintillation and Penning yields, the introduced system instantiates the concept of `Penning-Fluorescent' TPC, enabling i) $\times 10$ increased track sharpness as compared to pure Xenon, ii) wavelength-shifted primary scintillation and iii) stability of charge amplification based on micro-pattern readouts. The anticipated reduction on the Fano factor could not be demonstrated due to the presence of systematic instrumental limitations in the energy resolution determination. Further research along this promising line will continue.

\section{Acknowledgments}

The author deeply acknowledges countless discussions with colleagues
of the RD51 and NEXT collaborations. Calculations of Penning transfer rates
were performed by Ozkan Sahin. The excellent technical support and service from the CERN detector workshop
and Servicio General de Apoyo a la Investigaci´on-SAI of the University of Zaragoza
was essential to our investigation.
This work was funded by  Spanish Ministry of Economy and Competitiveness (MINECO)
under grants Consolider-Ingenio 2010 CSD2008-0037 (CUP) and CSD2007-00042 (CPAN),
FPA2011-24058, FPA2013-41085, FPA2008-03456, FPA2009-13697, and
by the T-REX Starting Grant ERC-2009-StG-240054 of the
IDEAS program of the 7th EU Framework Program. Part of these grants are funded
by the European Regional Development Plan (ERDF/FEDER).

%

\end{document}